\documentclass{article}
\usepackage{graphicx}

\begin{document}

\section*{The Shape of Space  after WMAP data}

\begin{center}
Jean-Pierre Luminet\\
 Laboratoire Univers et Th\'eories, CNRS-UMR 8102,
Observatoire de Paris, F--92195 Meudon c\'edex, France.

\end{center}

\subsection*{Abstract}

What is the shape of space is a long-standing question in cosmology.  
In this talk I review recent advances in cosmic topology since it has entered 
 a new era of experimental tests. High redshift surveys of 
astronomical sources and 
accurate maps of the Cosmic Microwave Background radiation (CMB)  are 
beginning 
to hint at the shape of the universe, or at least to limit 
the wide 
range of possibilities. Among those possibilites are  surprising 
``wrap around" 
universe models in which space, whatever its curvature, may be 
smaller than 
the observable universe and generate topological lensing effects on a 
detectable 
cosmic scale. 
In particular, the recent analysis of CMB
data provided by the WMAP satellite suggest a finite universe with 
the 
topology of the Poincar\'{e} dodecahedral
spherical space.  Such a  model of a ``small  universe", the  volume 
of which would represent only about 80{\nobreakspace}\% the volume 
of the observable universe, offers an observational signature 
in the form of a predictable topological lens effect on one hand, and 
rises new issues on the early universe physics on the other hand. 

\subsection*{The shape of the Universe}

The problem of the global shape of the universe  can be decomposed 
into three intertwined questions. \\
First, what is the space curvature ? In homogeneous isotropic models 
of relativistic cosmology, there are only three possible answers. 
Three-dimensional space sections of spacetime may have zero curvature 
on the average  -- in such a case, two parallel lines keep a constant 
space separation and never meet, as in usual Euclidean space, 
sometimes called ``flat  space". Or space sections can be negatively 
curved, 
such as two any parallels diverge and never meet (such a space is the 
three-dimensional 
analogue of the Lobachevsky hyperbolic plane). Eventually, they can 
be 
positively curved, in which case all parallels reconverge and cross 
again (like on the two--dimensional surface of a sphere).   \\
The property for physical space to correspond to one of these three 
possibilities depends on the way the total energy density of the 
Universe may counterbalance  the kinetic energy of the expanding 
space.
The normalized density parameter 
\ensuremath{\Omega}$_{0}$, defined as the ratio of the actual density 
to the critical value that an Euclidean space would require,  
characterizes the present-day contents (matter and all forms of 
energy) of the  
Universe.  If \ensuremath{\Omega}$_{0}$ is greater than 1, then space 
curvature is 
  positive and geometry  is spherical; if \ensuremath{\Omega}$_{0}$ 
is 
smaller than 1 the 
curvature is negative and geometry is hyperbolic; eventually 
\ensuremath{\Omega}$_{0}$ is 
strictly equal to  1 and space is Euclidean.  \\

The second question about the shape of the Universe is to know 
whether space is finite or infinite -- equivalent to know whether 
space 
contains a finite or an infinite amount of matter--energy, since the 
usual assumption of homogeneity implies a uniform distribution of 
matter and 
energy through space.  From a purely geometrical point of view, all 
positively curved spaces (called 
spherical spaces whatever their topology) are finite, but the 
converse is not true : flat (Euclidean) or negatively curved 
(hyperbolic) 
spaces can have finite or infinite volumes, depending on their degree 
of connectedness (Ellis, 1971{\nobreakspace}; Lachi\`{e}ze-Rey \& 
Luminet, 1995). For instance,  in a flat space with cubic torus 
topology,  as soon as a particle or a light ray ``exits" a given face 
of the 
fundamental cube, it  ``re-enters"  from the opposite face, so that  
space is finite, 
although without a boundary.   

From an observable point of view, it is necessary to distinguish 
between 
the ``observable universe", which is the interior of a sphere 
centered 
on the observer and whose radius is that of the cosmological horizon 
(roughly the radius of the last scattering surface), and the physical 
space. Again there are only three logical possiblities. First, the physical 
space is infinite  -- like for instance the simply-connected Euclidean space.
In this case, the observable universe is an infinitesimal 
patch 
of the full universe and, although it has long been the preferred 
model of many cosmologists, this is not a testable hypothesis. 
Second, physical space is finite (e.g. an hypersphere or a closed 
multiconnected space), but greater than the observable space.  In 
that case, one easily figures 
out that if physical space is much greater that the observable one, 
no signature of its finitude will show in the observable data.  
But if space is not too large, or if space is not globally 
homogeneous (as is permitted in many space models with multiconnected 
topology) and if the observer occupies a special position, some 
imprints of the space finitude could be observable. Third, 
physical space is smaller than the observable universe. Such an 
apparently odd possibility is due to the fact that space can be multiconnected and 
have a small volume. 
There a lot of geometrical possibilites, whatever the curvature of 
space. As it is well-known, such ``small universe" models may generate multiple 
images of light sources, in such a way that the hypothesis can  be 
tested by astronomical observations.    \\

The third question about the shape of the Universe  deals with its 
global topological properties (see Luminet, 2001 for a non-technical 
book about all the aspects of topology and its applications to 
cosmology). It is interesting to point out that none of these 
global properties is given by Einstein's field equations, since they are 
partial differential equations describing only the local, metric 
structure of spacetime (Friedmann, 1924). The present-day topology 
and curvature of space take likely their origin in the early quantum 
conditions of the Universe, which also governed its time evolution. 
The topological classification of homogeneous Riemannian 3-D spaces 
has made considerable progress during the last century. There are 18 
Euclidean spaceforms (for a full description, see Riazuelo et al., 
2004), a countable infinity of spherical spaceforms (see Gausmann et 
al, 2001) and a non-countable infinity of hyperbolic spaceforms (see 
Weeks, 1999.)  \\

\subsection*{Cosmic Crystallography}

The topology and the curvature of space  can be studied by using 
specific astronomical observations. For instance, from Einstein's 
field equations, the space curvature can be deduced from the 
experimental values of the total energy density and of the expansion 
rate. If the Universe was finite and small enough, we should be able 
to see ``all around" it, 
because the photons might have crossed it once or more times. In such a 
case, any observer might identify multiple images of a same light
source, although distributed in different directions of the sky and 
at various redshifts,  or 
to 
detect specific statistical properties in the apparent distribution 
of faraway sources such as galaxy clusters. To do this,  methods of 
``cosmic crystallography" have been devised (Lehoucq et al., 1996, 
1999, 2000), and extensively studied by the Brazilian school of 
cosmic topology (Gomero et al., 2000, 2001a, 2002a,b, 2003; Fagundes 
\& Gausmann, 1999) ; see also Marecki et al. (2005).

Basically, cosmic crystallography looks at the 3-dimensional apparent 
distribution of high redshift sources (e.g. galaxy clusters, quasars) 
in order to 
discover repeating patterns in the universal covering space, much like the 
repeating
patterns of atoms observed in a crystal.  ``Pair 
Separation Histograms" (PSH) are in most cases able to detect a 
multiconnected 
topology of space, in the form of sharp spikes standing out above the 
noise 
distribution that is expected in the simply-connected case. 
Figures 1-3 visualize the ``topological lens effect"  generated by a 
multiconnected shape of space, and the way the topology can be 
determined by the PSH method.

\begin{figure}[htbp]
\begin{center}
\includegraphics[scale=4.00]{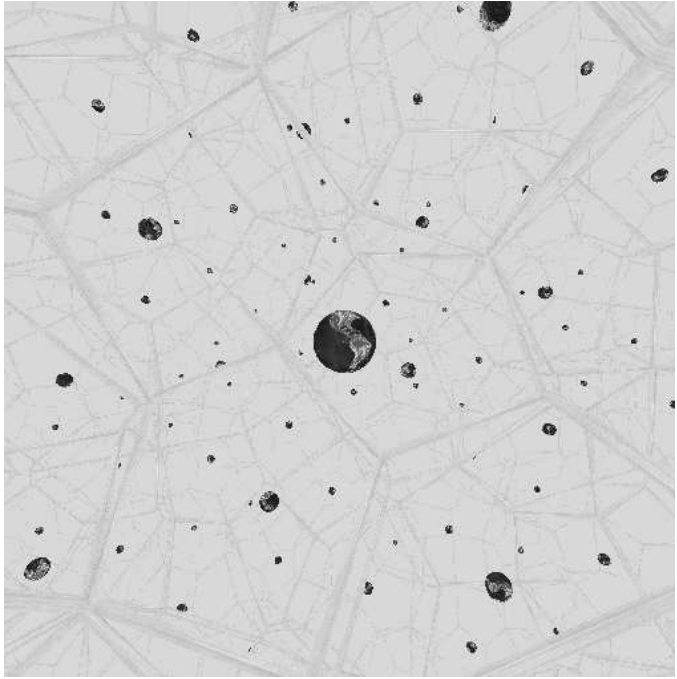}
\caption{{\it In a multi-connected Universe, the physical space is 
identified to a fundamental polyhedron, the duplicate images of which 
form the 
observable universe.
Representing the structure of apparent space is equivalent to 
representing its ``crystalline" structure, each cell of which is a 
duplicate of 
the fundamental polyhedron.
Here is depicted the closed hyperbolic Weeks space (only one 
celestial 
object is depicted, namely the Earth). As viewed from 
inside, it gives the illusion of a cellular space, tiled par 
polyhedra 
distorted with optical illusions (courtesy Jeffrey Weeks).}
}
\end{center}
\end{figure}

\begin{figure}[htbp]
\begin{center}
\includegraphics[scale=0.50]{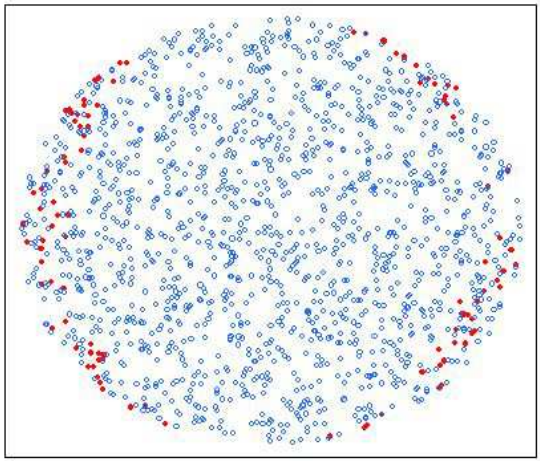}
\caption{{\it Sky map simulation in hypertorus flat space (left). The 
fundamental polyhedron is a cube with length = 60 \% the horizon size 
and contains 100 ``original" 
sources (dark dots). 
One observes 1939 topological images (light dots). }
}
\end{center}
\end{figure}

\begin{figure}[htbp]
\begin{center}
\includegraphics[scale=0.50]{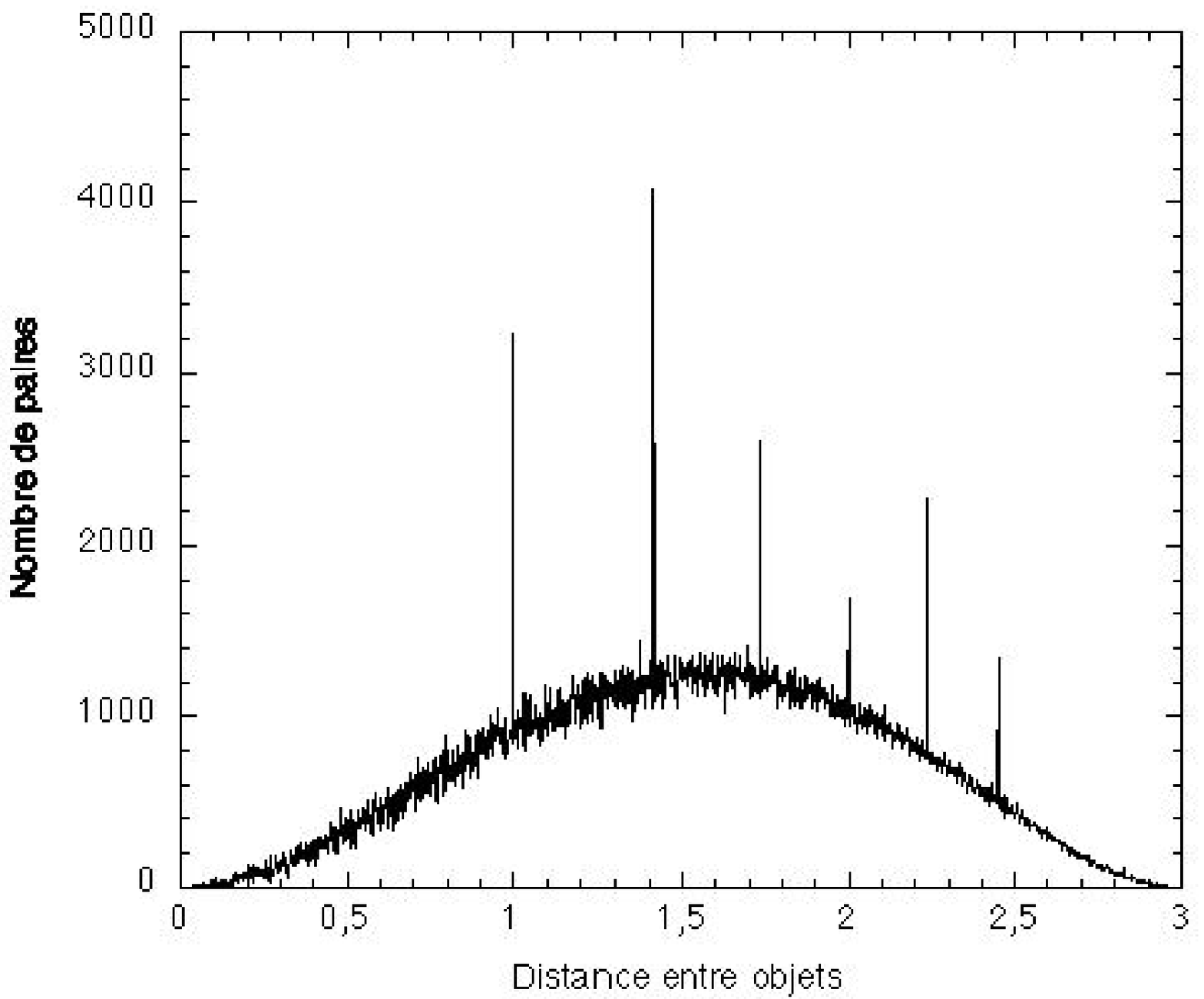}
\caption{{\it The Pair Separation 
Histogram corresponding to Figure 2 exhibits spikes which stand out at values and with 
amplitudes 
depending on the topological
properties of space.}
}
\end{center}
\end{figure}

However it was shown (Lehoucq et al., 2000; Gomero 
et al., 2002b ) that PSH may provide a topological signal only when the 
holonomy group of space has 
Clifford translations, a property which excludes all hyperbolic 
spaces. 

\subsection*{Spherical Lensing}

In the first investigations of cosmic topology, the search for the 
shape of space had focused on big 
bang models with flat or negatively curved spatial sections. Since 
1999 however, a combination of astronomical (type I supernovae) and 
cosmological (temperature anisotropies of the CMB) observations 
suggest that the expansion of the universe is 
accelerating, and constrain the value of space curvature in a range 
which marginally 
favors a positively curved (i.e. spherical) model. As a consequence, 
spherical  spaceforms have come back to the forefront of cosmology. \\
Gausmann et al. (2001) have investigated the full properties of 
spherical universes. The simplest case is the celebrated hypersphere, 
which is finite yet with no boundary.  Actually there are an infinite number of spherical spaceforms, 
including lens spaces, prism spaces and polyhedral spaces. Gausmann 
et al. (2001) gave the
 construction and complete classification of 
such spaces, and discussed which topologies were 
likely to be detectable by crystallographic methods. They predicted 
the shapes of the pair separation histograms and they checked their 
predictions by computer 
simulations.  \\
In addition, Weeks et al. (2003) and Gomero et al. (2001b) proved that the spherical topologies 
would be more easily detectable 
observationally than hyperbolic or flat ones.  The reason is that, no 
matter how close space is to perfect 
flatness, only a  finite number of spherical shapes are excluded by 
observational 
constraints. 
Due to the special structure of spherical spaces, topological 
imprints would be potentially detectable within the observable 
universe. Thus 
cosmologists are taking a renewed interest in spherical spaces as 
possible models for 
the  physical universe. \\

\subsection*{The Universe as a drumhead}

The main limitation of cosmic crystallography is that the presently 
available catalogs of observed 
sources at high redshift are not complete enough to perform 
convincing tests (Luminet and Roukema, 1998).  \\
Fortunately, the topology of a small Universe may also be detected 
through its effects on such a  ``Rosetta stone" of cosmology as is 
the CMB 
fossil radiation (Levin, 2002{\nobreakspace}; Riazuelo et al., 
2004a).\\
If you sprinkle fine sand uniformly over a drumhead and then make it 
vibrate, the grains of sand will collect in characteristic spots and 
figures, 
called  Chladni patterns. These patterns reveal much information 
about the  size and the shape of the drum and the elasticity of its 
membrane. 
In particular, the distribution of spots depends not only on the way 
the drum vibrated initially but also on the global shape of the drum, 
because the waves will be reflected differently according to whether 
the edge of the drumhead is a circle, an ellipse, a square, or some 
other shape.\\ 
In cosmology, the early Universe was crossed by real acoustic waves 
generated soon after the big bang. Such vibrations left their imprints 
380 000 years later as tiny density fluctuations in the primordial 
plasma. Hot and 
cold spots in the present-day 2.7 K CMB radiation reveal those 
density fluctuations. Thus the CMB temperature fluctuations look like 
Chladni 
patterns resulting from a  complicated three-dimensional drumhead 
that vibrated for 380 
000 years. 
They yield a wealth of information about the physical conditions that 
prevailed in the early Universe, as well as present geometrical 
properties like space curvature and topology. More precisely, density 
fluctuations 
may be expressed as combinations of the vibrational modes of space, 
just as 
the vibration of a drumhead may be expressed as a combination of the 
drumhead's harmonics. 
The shape of space can be heard 
in a unique way. Lehoucq et al. (2002) calculated the 
harmonics (the so-called ``eigenmodes of the Laplace operator") for 
most of the spherical topologies, and Riazuelo et al. (2004b) did the 
same for all 
18 Euclidean spaces. Then, starting from a set of initial conditions
fixing how the universe originally vibrated (the so-called 
Harrison-Zeldovich spectrum), 
they evolved the harmonics forward in time to simulate realistic CMB 
maps for a number of flat and spherical topologies (Uzan et al., 2003a). 

\begin{figure}[htbp]
\begin{center}
\includegraphics[scale=1.0]{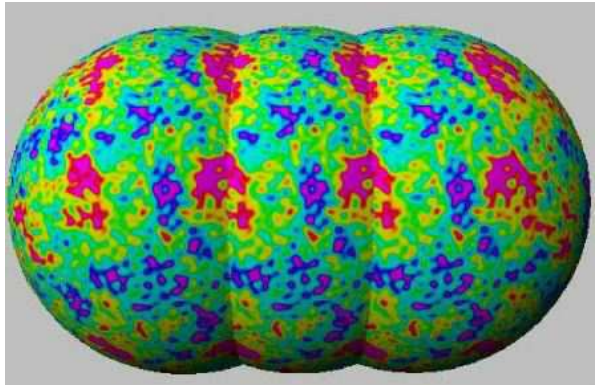}
\caption{{\it A multiconnected topology translates into the fact that 
any object in space may possess several copies of itself in the 
observable Universe. For 
an extended object like the region of emission of the CMB radiation 
we observe (the so-called last scattering surface) it can happen that 
it 
intersects with itself 
along pairs of circles. In this case, this is equivalent to say that 
an observer (located at the center of the last scattering surface) 
will 
see the same region of the Universe from different directions. As a 
consequence, 
the temperature 
fluctuations will match along the intersection of the last scattering 
surface with itself, as illustrated in the above figure. 
This CMB map is simulated for a multiconnected flat space -- namely a 
cubic hypertorus whose length is 3.17 times smaller than the diameter 
of the last scattering surface. Only two duplicates are depicted.}}
\end{center}
\end{figure}

\subsection*{Primordial fluctuations}

\begin{figure}[htbp]
\begin{center}
\includegraphics[bb = 20 20 592 306, scale=0.60]{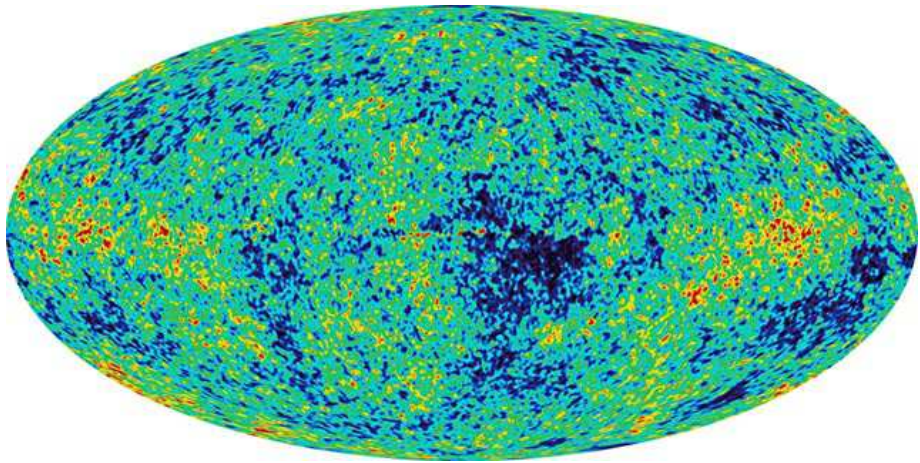}
\caption{{\it Map of temperature anisotropies of 
CMB as observed by WMAP telescope. WMAP Homepage{\nobreakspace}: 
http://map.gsfc.nasa.gov}}
\end{center}
\end{figure}

The ``concordance  model"  of cosmology describes the Universe 
as a flat infinite space in eternal expansion, accelerated under the 
effect  of a repulsive ``dark energy". The data collected by the NASA 
satellite WMAP (Bennett et al, 2003{\nobreakspace}; Spergel et al., 
2003) has recently produced a high 
resolution map of the CMB which showed the seeds of galaxies and 
galaxy clusters (figure 
5) and  allowed to check the validity of the dynamic 
part of the  expansion model. However, combined with other 
astronomical 
data  (Tonry et al., 2003), they suggest a  value of the density 
parameter \ensuremath{\Omega}$_{0}$ = 1.02 
\ensuremath{\pm} 0.02 at the $1\sigma$ level. The result is marginally compatible with 
strictly flat space sections. Improved measurements could indeed 
lower the
value of \ensuremath{\Omega}$_{0}$ closer to the critical value 1,  
or even below to the hyperbolic case. Presently however, taken at their face value, WMAP data 
favor a positively curved space, necessarily of finite volume since 
all 
spherical spaceforms possess this property.
This provides (provisory) answers to the first two questions stated 
above. \\

Now what about space topology{\nobreakspace}? There is an intriguing 
feature in WMAP data, already present in previous COBE mearurements 
(Hinshaw et al., 1996), although at a level of precision that was not 
significant enough 
to draw firm conclusions. The power spectrum of temperature 
anisotropies  (figure 
6) exhibits a set of ``acoustic" peaks when anisotropy is measured on small and 
mean  scales (i.e. concerning regions of the sky of relatively 
modest 
size). 
These peaks are remarkably consistent with the infinite flat space 
hypothesis. However, at large angular 
scale (for CMB spots typically separated by more than 60 \ensuremath{^\circ}),
there is a strong loss of power which deviates 
significantly from the 
predictions of the concordance model.  Thus it is necessary to look 
for an alternative.  

\begin{figure}[htbp]
\begin{center}
\includegraphics[width=4.0in]{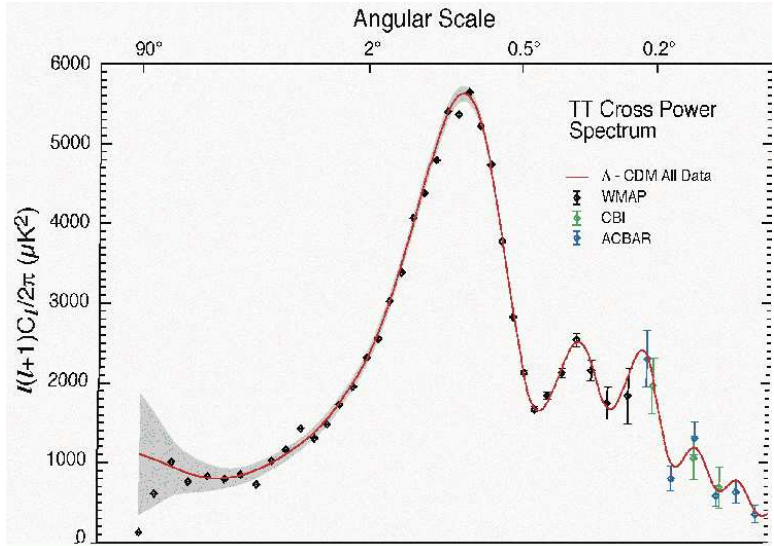}
\caption{{\it The CMB power spectrum depicts the minute 
temperature differences on the last scattering surface, depending on 
the angle of view. It shows a series 
of peaks corresponding to small angular separations (the position and 
amplitude of the main peak allows us to  measure space curvature), 
but at  larger angular scales, peaks disappear. According to the 
predictions 
of  the concordance model (continuous curve), at such scales the 
power 
spectrum  should follow the so-called ``Sachs-Wolfe plateau". 
However, WMAP measurements in this region (black diamonds) fall well 
below the plateau for the quadrupole and the octopole 
 moments (first two diamonds on the left). While the flat infinite 
space model cannot explain this feature, multiconnected space 
models  with a ``well-proportioned" topology are remarkably 
consistent with 
such data.  WMAP Homepage{\nobreakspace}: 
http://map.gsfc.nasa.gov}}
\end{center}
\end{figure}

CMB temperature anisotropies essentially result from density 
fluctuations of the primordial Universe : a photon coming from a 
denser region 
will  loose a fraction of its energy to compete against gravity, and 
will reach  us cooler. On the contrary, photons emitted from less 
dense 
regions will be received hotter. The density fluctuations result from 
the 
superposition of  acoustic waves which propagated in the  primordial 
plasma.
Riazuelo et al. (2004a) have developed complex theoretical 
models to reproduce the amplitude of such fluctuations, which can be 
considered as vibrations of the Universe itself. 
In particular,they simulated high resolution CMB maps for various 
space topologies (Riazuelo et al., 2004b ; Uzan et al., 2003a) and were able to compare their results with 
real WMAP  data. Depending on the underlying topology, the 
distribution of 
the fluctuations  differs. For instance, in an infinite flat space, 
all 
wavelengths are allowed, and fluctuations must be present at all 
scales.

\subsection*{Cosmic Harmonics}

The CMB temperature fluctuations can be decomposed into a sum of {\it 
spherical harmonics}, much like 
the sound produced by a music instrument may be decomposed into 
ordinary harmonics.  The ``fundamental" fixes the height of the note 
(as for 
instance a 440 hertz acoustic frequency fixes the \textit{A}  of the 
pitch), whereas the 
relative  amplitudes of each harmonics determine the tone quality 
(such as the \textit{A}  played by a piano differs from the 
\textit{A} 
played by a harpsichord). Concerning the relic radiation, the 
relative  amplitudes of each spherical harmonics determine the power 
spectrum, which is a signature of the geometry of space and of the 
physical  conditions 
which prevailed at the time of CMB  emission.\\

The first  observable harmonics is the quadrupole (whose wavenumer is 
$ \ell =2$). WMAP has observed a value of the quadrupole 7 times 
weaker 
than expected in a flat infinite Universe. The  probability that such 
a discrepancy occurs by chance has been estimated to  0.2 \% only. 
The octopole (whose wavenumber is $\ell =3$) is also weaker 
(72{\nobreakspace}\%  of  the expected value).  For larger 
wavenumbers up to $\ell =900$ 
(which correspond to temperature fluctuations at small angular 
scales),  observations are remarkably consistent with the standard 
cosmological 
model.\\

The unusually low quadrupole value means that  
long wavelengths are missing. Some cosmologists have proposed to 
explain the anomaly by still unknown physical laws of the early 
universe (Tsujikawa 
et al., 2003). 
A more natural explanation may be because space is not big enough to 
sustain long wavelengths. Such a situation may be compared to a 
vibrating string 
fixed at its two extremities, for which the maximum wavelength of an 
oscillation is twice the string length. On the contrary, in an 
infinite flat space, 
all the wavelengths are allowed, and 
fluctuations must be present at all scales.
 Thus this geometrical  explanation relies on a model of finite 
space whose size \textit{smaller} than the observable universe 
constrains the observable wavelengths below a maximum value.

\subsection*{Well-proportioned Spaces}

Such a property has been known for a long time, and was used to 
constrain the topology from COBE observations (Sokolov, 
1993{\nobreakspace}; Starobinsky, 
1993). Preliminary oversimplified  analyses (Stevens et al., 
1993{\nobreakspace}; de Oliveira-Costa \& Smoot, 
1995) suggested that any multi-connected topology in which 
space was finite in at least  one space direction had the effect of 
lowering the 
power spectrum at large wavelengths. Weeks et al. (2004) 
reexamined the question and showed that indeed,  some finite 
multiconnected topologies do lower the large--scale fluctuations
whereas others 
may elevate them.  In fact, the long wavelengths modes tend to be 
relatively lowered only in a special family of closed multiconnected 
spaces called ``well-proportioned". Generally, 
among spaces whose characteristic lengths 
are comparable with the radius of the last scattering surface 
$R_{lss}$ 
(a necessary condition for the topology to have an observable influence 
on the power spectrum), spaces with all dimensions of similar 
magnitude lower the quadrupole more heavily than the rest of the power 
spectrum. 
As soon as one of the characteristic lengths becomes significantly 
smaller or greater than the other two, the quadrupole is boosted in a 
way not compatible with WMAP data. The property was proved 
geometrically 
(Weeks et al., 2004), and checked out by numerical simulations 
(Riazuelo et al., 2004a).  In the case of flat 
tori, they have varied their proportions and shown that a cubic torus 
lowers the quadrupole whereas an oblate or a prolate torus increase 
the quadrupole. They have also studied 
spherical spaces and shown that polyhedric spaces suppress the 
quadrupole whereas high order lens spaces (strongly anisotropic) 
boost the quadrupole. Thus, well-proportioned 
spaces match the WMAP data much better than the infinite flat space 
model.

\subsection*{The Poincar\'{e} Dodecahedral Space}

Among the family of well-proportioned spaces, the best fit to 
the observed power spectrum is the {\it 
Poincar\'{e} Dodecahedral Space} (hereafter PDS) (Luminet et al., 
2003). \\
PDS  may be represented by a dodecahedron (a 
regular polyhedron with  12 pentagonal faces) whose opposite faces 
are glued after a 36\ensuremath{^\circ} twist (figure 7). Such a space is 
positively curved, and 
is a  multiconnected variant of the simply-connected hypersphere 
$S^{3}$,  with a volume 120 times smaller. 
A rocket going out of the dodecahedron by crossing a given face 
immediately re-enters by the opposite face. Propagation of light rays 
is  such that any observer whose line-of-sight intercepts one face 
has the 
illusion to see inside a copy of his own dodecahedron (figure 
8).

\begin{figure}[htbp]
\begin{center}
\includegraphics[scale=0.30]{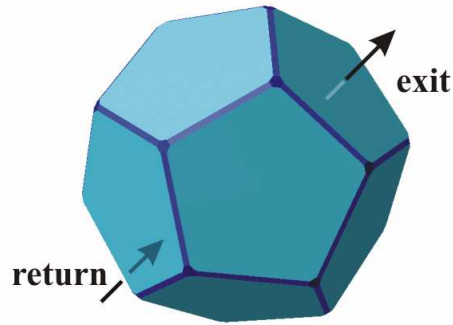}
\caption{{\it Poincar\'{e} Dodecahedral Space can be described as the 
interior of 
a dodecahedron such that when one goes out from a pentagonal face, 
one comes back 
immediately inside the space from the opposite face, after a 
36\ensuremath{^\circ} rotation.
Such a space is finite, although without edges or boundaries, so 
that one can indefinitely travel within it.}}
\end{center}
\end{figure}

\begin{figure}[htbp]
\begin{center}
\includegraphics[scale=0.30]{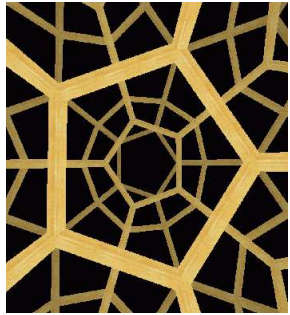}
\caption{{\it  View from inside PDS 
perpendicularly to one 
pentagonal face. In such a direction, ten 
dodecahedra tile together with a 1/10th turn to tessellate the 
universal covering space $S^3$. Since the dodecahedron has 12 faces, 
120 
dodecahedra are necessary to tessellate the full hypersphere. Thus, an observer has the 
illusion to live in a space 120 times vaster, made of tiled 
doecahedra which duplicate like 
in a mirror hall (courtesy Jeffrey Weeks).}}
\end{center}
\end{figure}

The associated  power spectrum, namely the repartition of 
fluctuations as a 
function of their wavelengths corresponding to PDS, 
strongly depends on the value of the mass-energy density 
parameter. Luminet et al. (2003) computed the CMB multipoles por 
$\ell = 2,3, 4$ and fitted the overall normalization factor to match 
the WMAP data at $\ell = 4$, and then examined their prediction for the 
quadrupole and the octopole as a function of $\Omega_{0}$. There is a small interval 
  of values within which the spectral fit is excellent, and in agreement 
with the value of the total density parameter deduced from WMAP data 
($1.02 \pm 0.02$).  The best fit is obtained for 
\ensuremath{\Omega}$_{0}$ = 1.016 (figure  9).
The result is quite remarkable because the Poincar\'{e} 
  space has no degree of freedom. By contrast, a 3-dimensional torus, 
  constructed by gluing together the opposite faces of a cube 
  and which constitutes a possible topology for a finite Euclidean 
space,  may be deformed into any parallelepiped : therefore its 
geometrical 
  construction depends on  6 degrees of freedom. 

\begin{figure}[htbp]
\begin{center}
\includegraphics[bb = 0 0 587 458, scale=0.48]{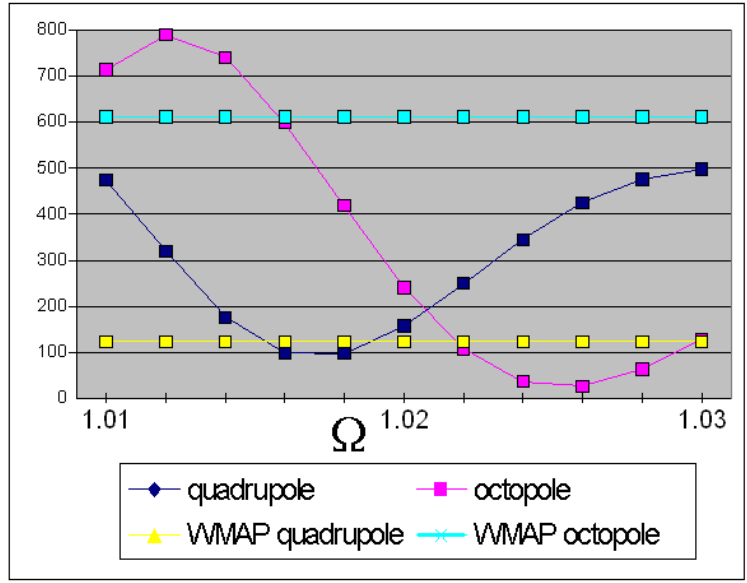}
\caption{{\it The values of the total mass-energy density parameter 
(assuming $\Omega_{m} = 0.28$) for 
which 
the  Poincar\'{e} Dodecahedral Space fits the WMAP observations. }
}
\end{center}
\end{figure}

The values of the matter density \ensuremath{\Omega}$_{m}$, of the dark 
energy density   
\ensuremath{\Omega}$_{\ensuremath{\lambda}}$ and of the expansion rate 
H$_{0}$ 
fix the radius of the last scattering surface $R_{lss}$ as well as 
the curvature radius of space $R_{c}$, thus dictate the possibility to 
detect the topology or not. For \ensuremath{\Omega}$_{m}$ = 0.28, 
\ensuremath{\Omega}$_{0}$ = 1.016 and H$_{0}$ = 
62 km/s/Mpc, $R_{lss}$ \ensuremath{\sim} 53 Gpc and $R_{c}$ = 2.63 
$R_{lss}$. It is to be noticed that the curvature radius $R_{c}$ is the 
same for the simply-connected universal covering space  $S^{3}$ and 
for the multiconnected PDS. Incidently, the numbers above show that, 
contrary to a current opinion, a cosmological model with 
\ensuremath{\Omega}$_{0}$ \ensuremath{\sim} 1.02 is far  
from being ``flat" (i.e. with $R_{c} = \infty$){\nobreakspace}! For the same curvature 
radius, PDS has a volume 120 times smaller than  S$^{3}$. Therefore, 
the smallest dimension of the fundamental dodecahedron is only 43 
Gpc, 
and its volume about 80\% the volume of the observable universe 
(namely the volume of the last scattering surface). This implies that some points of 
the last scattering surface will have several copies. Such a lens effect 
is purely attributable to topology and can be precisely calculated in 
the framework of the PDS model. It provides a definite 
signature of PDS topology, whereas the shape of the power 
spectrum gives only a hint for a small, well-proportioned 
universe model.  

To resume, the Poincar\'{e} Dodecahedral Space accounts for the low 
value of 
the  quadrupole as observed by WMAP in the fluctuation spectrum, and 
provides a  good value of the octopole. To be confirmed, the PDS 
model, which has been popularized as the ``soccerball universe model", must 
satisfy two experimental tests :\\
1) A finer analysis of WMAP data, or new data from the future 
European satellite ``Planck Surveyor" (scheduled 2007), will be able 
to determine the 
value of the energy density parameter with a precision of 
1{\nobreakspace}\%. A value lower 
than 1.01 will discard the Poincar\'{e} space as a model for 
cosmic space, in the sense that the size of the corresponding 
dodecahedron 
would become greater than the observable universe and would not leave
any 
observable imprint on the CMB, whereas a value greater than 1.01 
would 
strengthen its cosmological pertinence.\\
2) If space has a non trivial topology, there must be particular 
correlations in the CMB, namely pairs of ``matched circles" along 
which temperature fluctuations should be the same (Cornish et al, 
1998). 
The PDS model predicts 6 pairs of 
antipodal circles with an angular radius less 
than 35\ensuremath{^\circ}. \\
Such circles have been searched in WMAP data by two different teams, 
using various statistical indicators and massive  
computer calculations. 
On the one hand, Cornish et al. (2004) claimed to have found no 
matched circles on angular 
sizes greater than 25\ensuremath{^\circ}, and thus rejected the PDS 
hypothesis. Moreover, they claimed that any reasonable topology 
smaller than the 
horizon was excluded. This is a wrong statement because they searched 
only for antipodal or nearly-antipodal matched circles. However 
Riazuelo et al. (2004b) have shown that for generic topologies 
(including the well-proportioned topologies which are good candidates 
for explaining the WMAP power spectrum), the matched circles are not 
back-to-back and space is not globally homogeneous, so that the 
positions of the matched circles depend on the observer's position in the 
fundamental polyhedron. The corresponding larger number of 
degrees of freedom for the circles search in the WMAP data generates a dramatic 
increase of the computer time, up to values which are out--of--reach 
of the present facilities.  \\
On the other hand, Roukema et al. (2004) performed the same analysis 
for 
smaller circles,  and found six pairs of matched circles distributed 
in a 
dodecahedral pattern, each circle on an angular size about 
11\ensuremath{^\circ}.
This implies  $\Omega_{0} = 1.010 \pm 0.001$ for  
$\Omega_m = 0.28 \pm 0.02$, values which are perfectly consistent 
with 
the PDS model. \\
It follows that the debate about the pertinence of PDS as the 
best fit to reproduce CMB observations is fully open. Since then, the 
properties of PDS 
have  been investigated in more details by various authors. 
Lachi\`eze-Rey (2004) found an analytical expression of the  
eigenmodes of PDS, whereas Aurich et al. (2005) computed numerically  
the first 10 521 eigenfunctions up to the $\ell = 155$ mode and 
also supported the PDS hypothesis for explaining WMAP data. 
Eventually, the second--year WMAP data, 
originally expected by February 2004 but delayed for at least one year 
due to unexpected surprises in the results, may soon bring additional 
support to a spherical multiconnected space model.

\subsection*{Consequences for the physics of the early universe}

Finite well--proportioned spaces, and specially the Poincar\'{e} 
dodecahedral spherical space, open something like a ``Pandora box" for 
the physics that prevailed in the early universe. The concordance 
model relies mostly on the hypothesis that the early universe 
underwent a phase of exponential expansion -- the celebrated 
``inflationary process". Even without mentioning topological 
subtleties, it is good to recall that inflation theory gets into some 
troubles. In the simplest inflationary models, space is supposed 
to have become immensely larger than the observable universe after its 
phase of exponential growth.
Therefore apositive curvature (i.e. \ensuremath{\Omega}$_{0}$ \texttt{>} 
1), even weak, implies a finite space and sets strong constraints on 
the number of e-foldings that took place during an inflation phase.  It is possible to build models 
of ``low scale" inflation where the inflationary phase is short and leads to a detectable 
space  curvature (Uzan et al., 2003b). It turns out that, if space is 
not flat, the possibility of a multiconnected topology is not in contradiction 
with the general idea of inflation, due a number of free and 
adjustable parameters in this kind of models. Yet, no convincing 
physical scenario has been proposed (see however Linde, 2003). \\
In most cosmological models, it is generally assumed that spatial 
homogeneity stays valid beyond the horizon scale. For instance, in 
the model of chaotic inflation (Linde et al., 1994{\nobreakspace}; Guth, 
2000), the universe could be very homogeneous but on scales much 
larger than the horizon scale. On this respect, the PDS model seems incompatible 
with chaotic inflation : it requires only one expanding bubble 
universe, of size sufficlently small to be entirely observable. In 
his seminal cosmological paper, Einstein (1917) had already 
emphasized that spatially closed universes had the advantage to 
eliminate boundary conditions (Wheeler, 1968). A small universe like 
the PDS or a well--proportioned one, in which the observer could have 
access to all the existing physical reality is still more advantageous (Ellis \& Schreiber, 
1986). It is the only type of model in which the astronomical future 
could be definitely predicted -- such as the return of Halley's 
comet --, because only in such universes the observer could access to all 
the data in order to perform such predictions.\\  
Maybe the most fundamental issue is to link the present--day topology 
of space  to a quantum origin, since classical general relativity does 
not allow for topological changes during the course of cosmic evolution.  
Theories of quantum gravity could allow to address the problem of 
a quantum origin of space topology.  For instance, in the approach of 
quantum cosmology, some simplified solutions of  Wheeler-de Witt 
equations show that the sum over all topologies involved in the 
calculation of the wavefunction of the universe is dominated by 
spaces with small volumes and multiconnected topologies (Carlip, 1993 ; 
e Costa \& Fagundes, 2001). 
In the approach of brane worlds (see Brax 2003 for a review), the 
extra--dimensions are often assumed to form a compact Calabi-Yau manifold{\nobreakspace}; 
in such a case, it would be strange that only the ordinary dimensions of 
our 3--brane would not be compact like the extra ones. \\
These are only heuristic indications on the way unified theories of 
gravity and quantum mechanics could ``favor"  multiconnected spaces. 
Whatsoever the fact that some particular multiconnected space models, such as 
PDS, may be refuted by future astronomical data, the question of cosmic 
topology will stay as a major question about the ultimate structure 
of our universe.

\subsection*{References}

Aurich W.,  Lustig S. \& Steiner F. (2005), astro-ph/0412569.\\

\noindent
Bennett, C. L. \textit{et al}. (2003), Astrophys. J. Suppl. 
Ser. \textbf{148,} 1--27.\\

\noindent
Brax P. \& Van de Bruck C. (2003) Class.Quant.Grav. \textbf{20}, 
R201-R232.\\

\noindent
Carlip S. (1993), Class. Quant. Grav. \textbf{10}, 207-218.\\

\noindent
Cornish, N., Spergel, D. \& Starkman, G. (1998), Class. Quant. 
Grav. \textbf{15,} 2657--2670.\\

\noindent
Cornish, N. J.,  Spergel, D. N., Starkman,  G. D. and Komatsu, 
E. (2004), 
Phys. Rev. Lett. \textbf{92,} 201302.\\

\noindent
de Oliveira-Costa A. \& Smoot G.F. (1995), Astrophys. J. 
\textbf{448}, 447.\\

\noindent
e Costa, S. S. \& Fagundes,  H. V. (2001),  
Gen.Rel.Grav. \textbf{33}, 1489-1494.\\

\noindent
Einstein, A. (1917), Preuss. Akad.Wiss. Berlin Sitzber\textit{.} 
142--152.\\

\noindent
Ellis, G. F. R. (1971), Gen. Rel. Grav. \textbf{2,} 7--21.\\

\noindent
Ellis, G. F. R. \& Schreiber, W. (1986), Phys. Lett. A \textbf{115,} 
97--107.\\

\noindent
Fagundes, H. V. \& Gausmann  E. (1999),  
Phys.Lett. \textbf{A261}, 235-239.\\

\noindent
Friedmann, A. (1924), Z. Phys. \textbf{21,} 326--332.\\

\noindent
Gausmann E., Lehoucq R., Luminet J.-P., Uzan J.-P. \& Weeks J. 
(2001), Class. Quant. Grav., \textbf{18}, 5155.\\

\noindent
Gausmann, E., Lehoucq,  R., Luminet,  J.-P.,  Uzan, J.-P.  and Weeks, 
J. (2001), Classical and Quantum 
Gravity, \textbf{18}, 1-32.\\

\noindent
Gomero, G.I. (2003), Class.Quant.Grav. \textbf{20}, 4775-4784.\\

\noindent
Gomero, G.I., Reboucas,  M.J., Teixeira, A.F.F. (2000),   
Phys.Lett. \textbf{A275}, 355-367 ; (2001a), 
Class.Quant.Grav. \textbf{18}, 1885-1906.\\

\noindent
Gomero, G.I.,  Reboucas, M.J.,  Tavakol, R. (2001b), Class.Quant.Grav. 
\textbf{18}, 4461-4476 ; (2002), 
Int.J.Mod.Phys. \textbf{A17}, 4261-4272.\\
ÊÊÊ
\noindent
Gomero, G.I., Teixeira, A.F.F., Reboucas, M.J.,  Bernui, A. (2002), 
Int.J.Mod.Phys. D11, 869-892.\\

\noindent
Guth, A. H. (2000), Phys. Rep. \textbf{333,} 555--574.\\

\noindent
Hinshaw G. et al. (1996), Astrophys. J. Lett. \textbf{464}, L17-20.\\

\noindent
Lachi\`{e}ze-Rey, M. (2004), Class.Quant.Grav. \textbf{21}, 2455-2464.\\

\noindent
Lachi\`{e}ze-Rey, M. \& Luminet, J. P. (1995), Phys. Rep. \textbf{254,} 
135--214.\\

\noindent
Lehoucq R., Lachi\`{e}ze-Rey M. \& Luminet J.P. (1996), Astron. 
Astrophys. \textbf{313}, 339-346.\\

\noindent
Lehoucq R., Luminet J.-P. \&  Uzan J.-P. (1999), Astron. Astrophys. 
\textbf{344}, 735.\\

\noindent
Lehoucq R., Uzan J.-P. \& Luminet J.-P. (2000), Astron. Astrophys. 
\textbf{363},  1.\\

\noindent
Lehoucq, R., Uzan, J.-P. \& Weeks, J. (2003), 
Class. Quant. Grav. \textbf{20}, 1529-1542.\\

\noindent
Lehoucq, R., Weeks,  J. , Uzan, J.-P., Gausmann, E. and Luminet, 
J.-P. (2002),
Classical and Quantum Gravity, \textbf{19}, 4683-4708. \\

\noindent
Levin J. (2002) Phys. Rep. \textbf{365}, 251-333.\\

\noindent
Linde A. (2003) JCAP \textbf{0305}, 002.\\

\noindent
Linde, A., Linde, D. \& Mezhlumian, A. (1994), Phys. Rev. D 
\textbf{49,} 1783--1826.\\

\noindent
Luminet, J.- P., {\it L'Univers chiffonn\'e}, Fayard, Paris, 2001. 
Portugese translation {\it Sentido e segredos do Universo}, Instituto 
Piaget, Lisboa, 2002.\\

\noindent
Luminet J.-P. \& Roukema B. (1998), Proc. Cargese 98 summer school 
{\it Cosmology : 
The Universe at Large Scale}, M. Lachieze-Rey (Ed.), Kluwer Ac. Pub., 
NATO ASI 970491.\\

\noindent
Luminet J.-P., Weeks J., Riazuelo A., Lehoucq R. \& Uzan J.-P. 
(2003), Nature \textbf{425}, 593-595.\\

\noindent
Marecki A., Roukema B., Bajtlik S. (2005), to appear in Astron.  \& 
Astrophys., astro-ph/0412181.\\ÊÊ

\noindent
Riazuelo, A., Uzan, J.-P., Lehoucq, R. \& Weeks, J. (2004a), 
Phys.Rev. D69, 103514.\\

\noindent
Riazuelo, A., Weeks, J., Uzan, J.-P., Lehoucq, R. \& Luminet, J.-P. 
(2004b), 
Phys Rev. D69, 103518.\\

\noindent
Roukema, B. F., Lew,  B., Cechowska, M.,  Marecki, A. \& Bajtlik, 
S. (2004), Astron. \& Astrophy. 423,
821.\\

\noindent
Sokolov I.Y. (1993), JETP Lett. \textbf{57}, 617.\\

\noindent
Spergel, D. N. \textit{et al.} (2003), Astrophys. J. Suppl. 
Ser\textit{.} \textbf{148,} 175--194.\\

\noindent
Starobinsky A.A.(1993), JETP Lett. \textbf{57}, 622.\\

\noindent
Stevens D., Scott D. \& Silk J. (1993), Phys. Rev. Lett. \textbf{71} 
20.\\

\noindent
Tonry \textit{et al.}, (2003) Astrophys.J. \textbf{594} 1-24.\\

\noindent
Tsujikawa, S.,Maartens, R. \& Brandenberger, R. H. (2003), e-print 
astro-ph/0308169.\\

\noindent
Uzan, J.-P., Riazuelo, A., Lehoucq, R. \& Weeks, J. (2003a), 
astro-ph/0303580, to appear in Phys. Rev. D.\\

\noindent
Uzan J.-P., Kirchner U., Ellis G.F.R. (2003b), Mon.Not.Roy.Astron.Soc. 
\textbf{344} L65.\\

\noindent
Weeks, J., {\it SnapPea: A computer program for creating and studying hyperbolic 3-manifolds},
available by anonymous ftp from http://geometrygames.org/SnapPea/\\

\noindent
Weeks, J., Lehoucq,  R. and Uzan, J.-P. (2003), Class.Quant.Grav. 20, 1529-1542. \\

\noindent
Weeks, J., Luminet, J.-P., Riazuelo, A. \&  Lehoucq, R. (2004),  
MNRAS,  352, 258-262. \\

\noindent
Wheeler, J. A. \textit{Einstein's Vision} (Springer, Berlin, 1968).\\

\end{document}